\documentstyle[12pt,amssym,aas2pp4]{article}
\def\Msolar{{M$_{\odot}$\,}}
\def\arcsec{{$^{\prime\prime}$}\,} 

\received{Feb 17, 1999 } 
\accepted{Aug 21, 1999 }

\slugcomment{To appear in ``The Astrophysical Journal''}

\begin{document}

\title{The Mass Function of Main Sequence Stars in NGC\,6397 from Near
IR and Optical High Resolution HST Observations\altaffilmark{1}}

\altaffiltext{1}{Based on observations with the NASA/ESA {\it Hubble
Space Telescope}, obtained at the Space Telescope Science Institute,
which is operated by AURA, Inc., under NASA contract NAS5-26555}

\author{Guido De Marchi, Francesco Paresce, and Luigi
Pulone\altaffilmark{2}} \affil{European Southern Observatory,
Karl-Schwarzschild Strasse 2, D--85748 Garching, Germany\\
demarchi@eso.org, fparesce@eso.org, lpulone@eso.org\\
\vskip 1cm
{\bf Accepted for publication in ``The Astrophysical Journal''}}
\authoremail{demarchi@eso.org, fparesce@eso.org, pulone@mporzio.astro.it} 
\altaffiltext{2}{On leave from the Osservatorio Astronomico di Roma, 
Rome, Italy}

\begin{abstract}
We have investigated the properties of the stellar mass function in the
globular cluster NGC\,6397 through the use of a large set of HST
observations. The latter include existing WFPC\,2 images in the $V$ and
$I$ bands, obtained at $\sim 4\farcm5$ and $10$\arcmin\, radial distances,
as well as a series of deep images in the $J$ and $H$ bands obtained
with the NIC\,2 and NIC\,3 cameras of the NICMOS instrument pointed,
respectively, to regions located $\sim 4\farcm5$ and $\sim 3\farcm2$
from the center. These observations span the region from $\sim 1$
to $\sim 3$ times the cluster's half-light radius ($r_{\rm hl} \simeq
3^{\prime}$), and have been subjected to the same, homogeneous data
processing so as to guarantee that the ensuing results could be
directly compared to one another. We have built color--magnitude
diagrams that we use to measure the luminosity function of main
sequence stars extending from just below the turn-off all the way down
to the Hydrogen burning limit. All luminosity functions derived in this
way show the same, consistent behavior in that they all increase with
decreasing luminosity up to a peak at $M_I \simeq 8.5$ or $M_H \simeq
7$ and then drop precipitously well before photometric incompleteness 
becomes significant. Within the observational uncertainties, at $M_I 
\simeq 12$ or $M_H \simeq 10.5$ ($\sim 0.09$\,\Msolar) the luminosity
functions are compatible with zero. The direct comparison of our NIC\,2
field with previous WFPC\,2 observations of the same area shows that
down to $M_H\simeq 11$ there are no more faint, red stars than those
already detected by the WFPC\,2, thus excluding a significant
population of faint, low-mass stars at the bottom of the main sequence.
By applying the best available mass--luminosity relation appropriate to
the metallicity of NGC\,6397 and consistent with our color--magnitude
diagrams to both the optical and IR data, we obtain a mass function
that shows a break in slope at $\sim 0.3$\,\Msolar. No single exponent
power-law distribution is compatible with these data, regardless of the
value of the exponent.  We find that a dynamical model of the cluster
can simultaneously reproduce the luminosity functions observed in the
core, at $\sim 3\farcm2$, $4\farcm5$, and $10^{\prime}$ away from the
center, as well as the surface brightness and velocity dispersion
profiles of red giant stars only if the model IMF rises as $m^{-1.6 \pm
0.2}$ in the range $0.8 - 0.3$\,\Msolar and then drops as $m^{0.2 \pm
0.1}$ below $\sim 0.3$\,\Msolar. Adopting a more physical log-normal
distribution for the IMF, all these data taken together imply a best
fit distribution with $m_c\simeq 0.3$ and $\sigma \simeq 1.8$.

\end{abstract}
 

\section{Introduction}

Among the Galactic globular clusters (GC), NGC\,6397 has so far been
the preferred target in the quest for the lowest mass stars that could
tell us much about the IMF of these systems and clarify whether or not
they can hide a substantial fraction of their total mass in the form of
very light objects. The close proximity of NGC\,6397 ($\sim 2$\,kpc)
coupled with its low metallicity ([Fe/H]$\sim -1.9$; Djorgovski 1993)
makes stars at the bottom of the main sequence (MS) relatively easy to
observe with the powerful cameras on board the HST, and many have
undertaken this challenge. Our original work on this subject (Paresce,
De~Marchi, \& Romaniello 1995) revealed for the first time a marked
deficiency of low-mass stars with respect to expectations based on a
power-law mass function (MF) increasing all the way to the
Hydrogen-burning limit (Fahlman et al. 1989). These findings were
questioned by Cool, Piotto, \& King 1996 and Chabrier \& M\'era (1997),
but have since been largely confirmed by independent measurements
(Mould et al. 1996; King et al. 1998) which all concur to suggest a MF
that flattens out below $\sim 0.3$\,\Msolar. We argue in this paper
that below $0.3$\,\Msolar the MF drops all the way to the detection limit
at the bottom of the stellar MS.

Although the results mentioned above seem to be consistent with each
other, and thus rather robust, there are at least two practical issues
that need to be carefully addressed.  First of all, it is fair to
wonder to which extent this deficiency of low-mass stars at the bottom
of the MS is physical and genuine and how much of it should rather be
attributed to some limitation inherent in the instrument or method of
data analysis used. In fact, the observations carried out so far are
confined to the optical domain ($V$ and $I$ bands), while low- and
very-low mass stars such as those near the H-burning limit are expected
to have a spectral energy distribution that peaks in the near-IR at
$\sim 1.2\,\mu$m (Allard \& Hauschildt 1995). If a large population of
very faint and redder stars were present in this cluster, they might
escape detection with the WFPC\,2. Secondly, the accurate determination
of a luminosity function (LF) from these observations requires a
reliable assessment of the contamination due to background unresolved
galaxies and to field stars which are rather numerous at the low
galactic latitude of NGC\,6397 ($b_{\rm II}\simeq -12$).  Attempts have
been made to account and correct for this contamination using both
color and proper motion information, but both methods are subject to
some uncertainties.

Finally, a third, more general question needs to be addressed
concerning the real meaning of these localized MF, namely whether the
observed flattening at low masses is a feature inherent in the global
MF (GMF) or, even, in the IMF of the cluster, or whether it simply
results from the internal dynamical modification of an originally
completely different mass distribution. Although Richer et al. (1991)
have proposed that, if measured near the half-mass radius, the present
day MF should closely resemble the IMF, no direct confirmation of this
hypothesis exists yet.

In this paper, we address all three issues above. The first two are
best attacked with the new NICMOS camera on board the HST, which we
have used to investigate the stellar population of NGC\,6397 precisely
at the wavelengths where low mass stars are expected to shine the most,
and to sample a control field located well outside of the cluster tidal
radius to provide an independent estimate of field object contamination. 
We address the third issue by modelling the dynamical state of
NGC\,6397 in Section\,5, where we put together all the LF currently
measured for this cluster at various radial distances and compare them
with the predictions of a multi-mass model that takes internal
dynamical evolution (mass segregation) into account.

\section{The Data}

\begin{figure}[h]
\plotfiddle{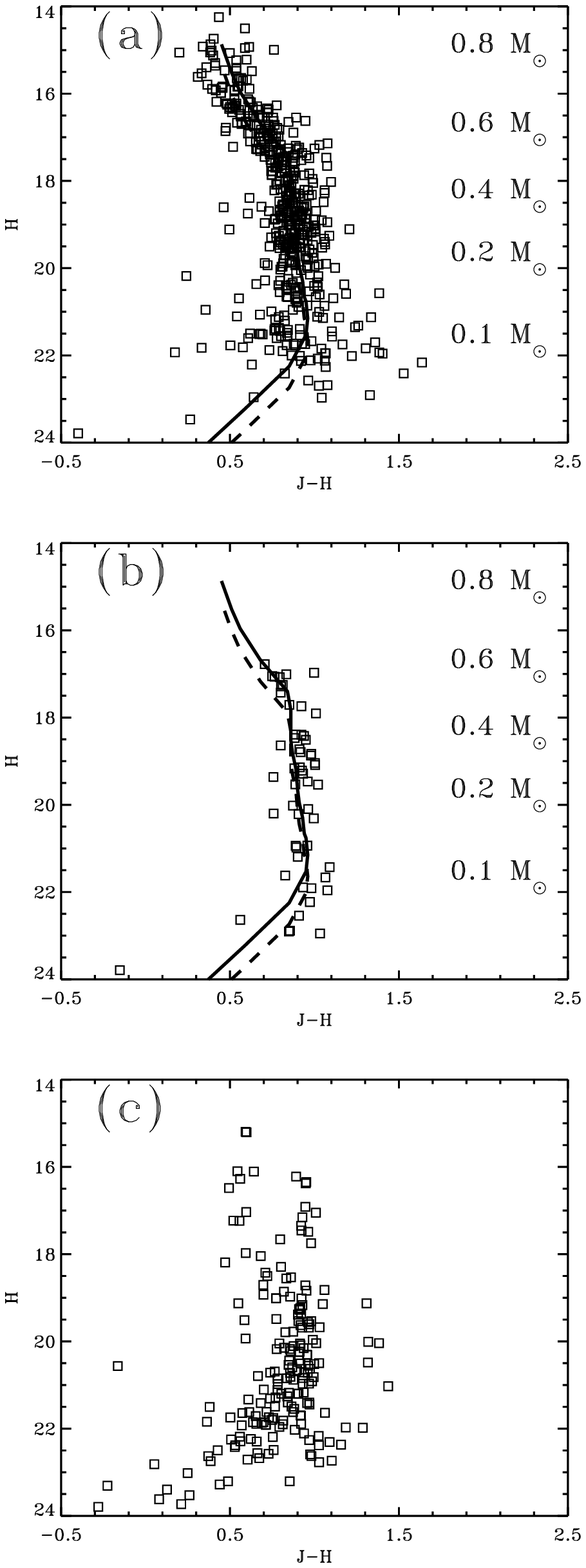}{20cm}{0}{80}{80}{-130}{0} 
\caption{IR Color-magnitude diagrams as measured in field\,F1 (a), F\,2
(b), and SKY (c). The magnitudes are given in the VEGAMAG system. See
the text for an explanation of the lines in panel (a) and (b).}
\end{figure}
 
The results of this paper are based on observations obtained by us with
both the NIC\,2 and NIC\,3 cameras of the NICMOS instrument on board
the HST (MacKenty et al. 1997). NIC\,2 was used on 1997, Sep 30 to
observe a region located $\sim 4\farcm6$\,NW of the center of NGC\,6397
(hereafter field F\,2) through the F110W and F160W broad band filters
with a total exposure time of 10,235\,s and 4,027\,s, respectively. The
field of view of the NIC\,2 camera ($\sim 19$\arcsec\,$\times
19$\arcsec) was placed over an area previously imaged with the WFPC\,2
(see Paresce et al. 1995). A field closer to the cluster center was
observed on 1998, June 17 with the NIC\,3 camera (during the second
``NIC\,3 Campaign''), when the $\sim 50$\arcsec$\times \,50$\arcsec FOV
was pointed to RA=17:40:54, DEC=--53:42:57 (J2000) corresponding to a
distance of $3\farcm2$ from the center (hereafter Field F\,1) and
images were taken for a total exposure duration of 2,303\,s in F110W
and of 703\,s in F160W. In units of the cluster's half-light radius
($r_{\rm hl} = 175$\arcsec Djorgovski 1993), our NIC\,2 and NIC\,3
fields are located, respectively, at $\sim 1.6\,r_{\rm hl}$ and $\sim
1.1\,r_{\rm hl}$. We also used the NIC\,3 camera to observe a
comparison field (hereafter SKY), located $2^o$\,NE of the cluster at
RA=17:47:52, DEC=--51:41:17 (J2000), to estimate the contamination due
to field stars and  correct for it in a statistical way. The filters
and exposure times are identical  to those used for the field F\,1.

The quality of all these data sets is excellent, with PSF full width at
half maximum (FWHM) of $\sim 0\farcs13$ for NIC\,2 and of $\sim
0\farcs19$ for NIC\,3 (achieved through the repositioning of HST's
secondary mirror operated during the NIC\,3 campaign). Images were
subjected to the standard STSDAS calibration (CALNICA) to remove the
instrumental signature (bias and flat field correction) and, in the
case of the NIC\,2 images, multiple frames of the same area and taken
through the same filter were combined to improve the statistics. The
average per-pixel count-rate of the background is of order
$0.03\,\pm\,0.015$\,count\,s$^{-1}$ in F110W and
$0.02\,\pm\,0.008$\,count\,s$^{-1}$ in F160W for the NIC\,2 data, and
$0.11\,\pm\,0.01$\,count\,s$^{-1}$ in F110W and
$0.12\,\pm\,0.01$\,count\,s$^{-1}$ in F160W for NIC\,3 images.  These
values, and particularly the standard deviation of the background, are
larger than those predicted with the NICMOS simulator, but still within
a factor of two of them.

The automated star detection routine {\it daophot.daofind} was applied
to the data, with a detection threshold conservatively set at
$5\,\sigma$ above the local average background level (the threshold was
set to $10\,\sigma$ for the NIC\,3 cluster field as it is more crowded
than the others, and the severe PSF under-sampling of NIC\,3 makes it
very difficult to detect and measure faint stars near brighter
objects).  We carefully examined by eye each individual object detected
by daofind and discarded saturated stars, a number of features (PSF
tendrils, noise spikes, etc.) that daofind had interpreted as stars, as
well as a few extended objects (with FWHM larger than twice that
typical of point sources). The number of well defined objects detected
in this way amounts to 517 in the F\,1 field, 58 in the F\,2 field,
and 119 in the SKY comparison field.  We should notice here that we
have used our knowledge of the quantum efficiency and filter
transmission of NICMOS in the F110W and F160W filters (see Pulone et
al. 1998) to select exposure times in these bands such that a typical
M-type dwarf in NGC\,6397 would be detected with a similar SNR in both
bands. And indeed, with an exposure time about $\sim 3$ times longer in
F110W than in F160W, all objects detected in one band are normally also
visible in the other.

Since crowding is not too severe in any of our images, stellar fluxes
were measured using the standard {\it digiphot.apphot} IRAF aperture
photometry routine, following the prescription of the ``core aperture
photometry'' technique described in De~Marchi et al. (1993).
Instrumental magnitudes were then calibrated and converted into the HST
magnitude system (STMAG) following the relation:

\begin{equation}
m_{ST}= -2.5 \log\left(\frac{c \, U}{\varepsilon \, t}\right)
\end{equation}

where $c$ is the number of counts measured for each star, $U$ the
inverse sensitivity of the instrumental setup (camera + filters),
$\varepsilon$ the encircled energy (i.e. the aperture correction), and
$t$ the total exposure time. The internal accuracy of our photometry
ranges from $<0.05$\,mag at $m_{160} \simeq 18$ to $\sim 0.5$\,mag at
$m_{160}\simeq 26$, although, particularly for the F110W observations
taken with NIC\,3, the highly variable pixel response function of the
camera can introduce statistical uncertainty of up to $\sim 10\%$
(Storrs 1998). Due to the current uncertainty affecting the photometric
zero points of the NIC\,3 camera (Nota et al. 1998), our absolute
photometry is accurate to within the $5\% - 10\%$ level.  In order to
make it easier to directly compare our results with those obtained from
the ground, we translate our measurements from the STMAG system into
the VEGAMAG photometric system of the HST, defined as one in which the
magnitude of Vega would be 0 in all bands and which is more similar to
the classical Johnson--Cousin ground-based system. We adopt the VEGAMAG
system in this paper and do so by subtracting the zero-point constants
of $2.3$\,mag and $3.7$\,mag from the values of the STMAG magnitudes in
the F110W and F160W bands, respectively. In the following, we refer to
these magnitudes as to $J$ and $H$, respectively.

With the magnitudes measured in this way we have produced the
color-magnitude diagrams (CMD) shown in Figure\,1 (panels a -- c).  The
MS of cluster stars is well defined in the F\,1 field (NIC\,3,
Figure\,1\,a), extending from $H \simeq 14.5$ to  $H \simeq 22$ where
it broadens due to increasing photometric errors. The MS of the stars
in the F\,2 field (Figure\,1\,b) is sparser and less well defined, but
it occupies the same region in the CMD as that of the F\,1 field. The
latter, however, appears consistently broader than the former at any
magnitude. This effect is likely to be the result of the non uniformity
of the sensitivity across the surface of the pixels of the NIC\,3
camera (Storrs 1998), coupled with the severe under-sampling of the
PSF. The expected ensuing uncertainty in the photometry is larger for
the F110W data (of order $11\,\%$) than for the F160W filter ($\sim
6\,\%$), and is in agreement with the observed MS broadening.

The lines in Figure\,1\,a and 1\,b represent the theoretical MS of a
10\,Gyr old population with the metallicity and distance appropriate to
NGC\,6397 as obtained using the models of Baraffe et al. (1997). We
have assumed the metal content $[Fe/H]=-1.9$ and extinction coefficient
$E(B-V)=0.18$ as given by Djorgovski (1993), but have let the distance
take on two different values, i.e. $(m-M)_{\rm O}=11.7$ from Djorgovski
(1993; solid line in Figure\,1) and $(m-M)_{\rm O}=12.2$ from Reid \&
Gizis (1998; dashed line). The theoretical sequences reproduce very
well our observations, and, particularly, the change of slope occurring
at $H \simeq 17.5$ and due to the change of the main source of opacity
in the atmosphere of the stars from H$^-$ to H$_2$. Because of the
uncertainty in the zero point of our photometry and of the broadening
of the MS due to the intra-pixel sensitivity variation of the NIC\,3
camera we cannot decide which of the two distance scales is more
appropriate, in that they are both consistent with our observations.
The mass values printed along the right-hand axes of Figure\,1 are
derived from the M-L relation of Baraffe et al. (1997) for a distance
modulus of $(m-M)_{\rm O}=12.2$ (Reid \& Gizis 1998).

\begin{figure}[h]
\plotone{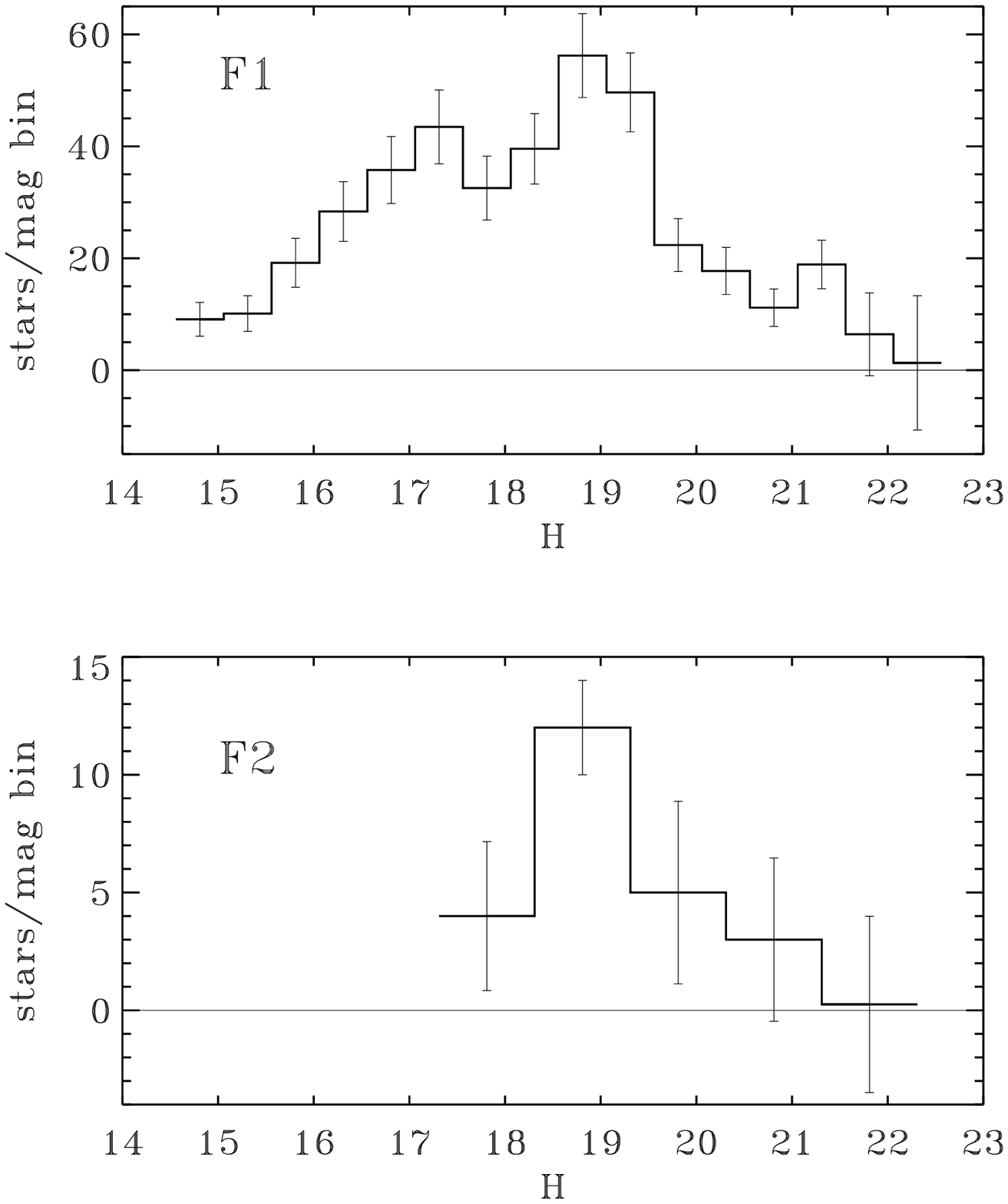}
\caption{IR luminosity functions as measured in field F\,1 (top
histogram) and F\,2 (bottom histogram). Error bars reflect the $\pm
1\,\sigma$ uncertainty of the Poisson statistics on the number counts
and the statistical uncertainty associated with the background
subtraction. Note, however, that we have associated a $2\,\sigma$ error
bar to the bins in which the net number of stars is $< 5$, as in these
cases the uncertainty can largely exceed the determined value and one
might prefer to use a $2\,\sigma$ upper limit instead. }
\end{figure}

Comparing the CMD of the F\,1 field with that of the SKY field
(Figure\,1\,c) immediately reveals that field stars (mostly disk
objects, owing to the low Galactic latitude of NGC\,6397, $b_{\rm
II}=-12$) occupy exactly the same location in the diagram as the
cluster objects.  It is thus not possible to assign cluster membership
to a star on the basis of its color alone.  This is precisely why we
have secured observations of the SKY field so as to be able to account
in a statistical way for the contamination due to field stars and
affecting our LF (see Section\,3).
 
A robust determination of the LF also requires proper correction for
the photometric incompleteness due to crowding. We have thus carried
out several standard artificial star tests by adding artificial stars
with magnitude values covering the whole range spanned by our
observations. For fields F\,2 and SKY we have generated 10 artificial
images per filter adding each time respectively 8 and 20 stars, and
have found that more than 90\,\% of the artificially added objects are
recovered at any magnitude level. The limited crowding and our
conservative choice of a $5\,\sigma$ detection threshold combine in
such a way that artificial stars are always retrieved unless they fall
onto another object. In field F\,1, however, crowding is more severe.
We ran 20 tests per filter adding 40 stars each time and found that
90\,\% of the artificially added objects are recovered at $H \simeq 19$
but the completeness gradually drops to $\sim 65\,\%$ at $H \simeq 22$
as shown in Table\,1.

\begin{table}
\caption{Table 1 -- Photometric completeness in the F\,1 field}
\begin{tabular}{cc}
\hline\hline
$H$ magnitude & completeness \\
\hline
15	& 100\,\%\\
16	& 99\,\% \\
17	& 97\,\% \\
18	& 94\,\% \\
19	& 90\,\% \\
20	& 88\,\% \\
21	& 75\,\% \\
22	& 67\,\% \\
\hline
\end{tabular}
\end{table}

To derive the LF of MS stars in the cluster we have counted the number
of objects as a function of the $H$ magnitude in the CMD corresponding
to fields F\,1 and F\,2.  We have determined in this way the number of
stars in each $0.5$ mag bin (1-mag bins were used for F\,2), which
is then multiplied by the incompleteness coefficient.  This latter step
needs only be performed for stars in field F\,1, as the photometric
incompleteness is negligible in fields F\,2 and SKY, where the Poisson
statistics of the counting process dominates the final uncertainty on
the LF. In order to account in a statistical way for field star
contamination, we then subtracted from each bin the number of objects
measured in the same magnitude bin of the CMD of field SKY (the area
covered by SKY is the same as F\,1 and seven times larger than F\,2).
The LF obtained in this way are shown in Figure\,2 in a linear scale
as a function of the observed $H$ band magnitude.

Although the LF in field F\,2 (bottom of Figure\,2) spans a narrower
magnitude range than that obtained with NIC\,3 in F\,1 (top histogram)
because of saturation of the few bright stars in field F\,2, within the
experimental errors the two LF agree very well with each other over
the common magnitude interval, once properly scaled. Indeed, although
the former LF is affected by large uncertainty, its shape is not
different from that measured in F\,1, with which it agrees well after
having been scaled vertically by a factor of $\sim 5$  to account for
the the larger area covered by F\,1. Both LF show a peak near $H\simeq
19$ followed by a clear drop all the way down to $H \simeq 22$, where
they become statistically consistent with zero. The error bars plotted
in Figure\,2 reflect the $\pm 1\,\sigma$ uncertainty obtained by
combining the Poisson statistics on the number counts with the
statistical uncertainty associated with the background subtraction. We
note here, however, that we have associated a $2\,\sigma$ error bar to
the bins in which the net number of stars is $< 5$, as in these cases
the uncertainty can largely exceed the determined value and one might
prefer to use a $2\,\sigma$ upper limit instead. Nevertheless, it is
noteworthy that the number of stars becomes compatible with being zero
at magnitude very close to that expected for the MS H-burning limit for
stars with the metallicity of NGC\,6397 (Baraffe et al.  1997).

\section{Comparing the Optical to the Infra-Red Data}

\begin{figure}[h]
\plottwo{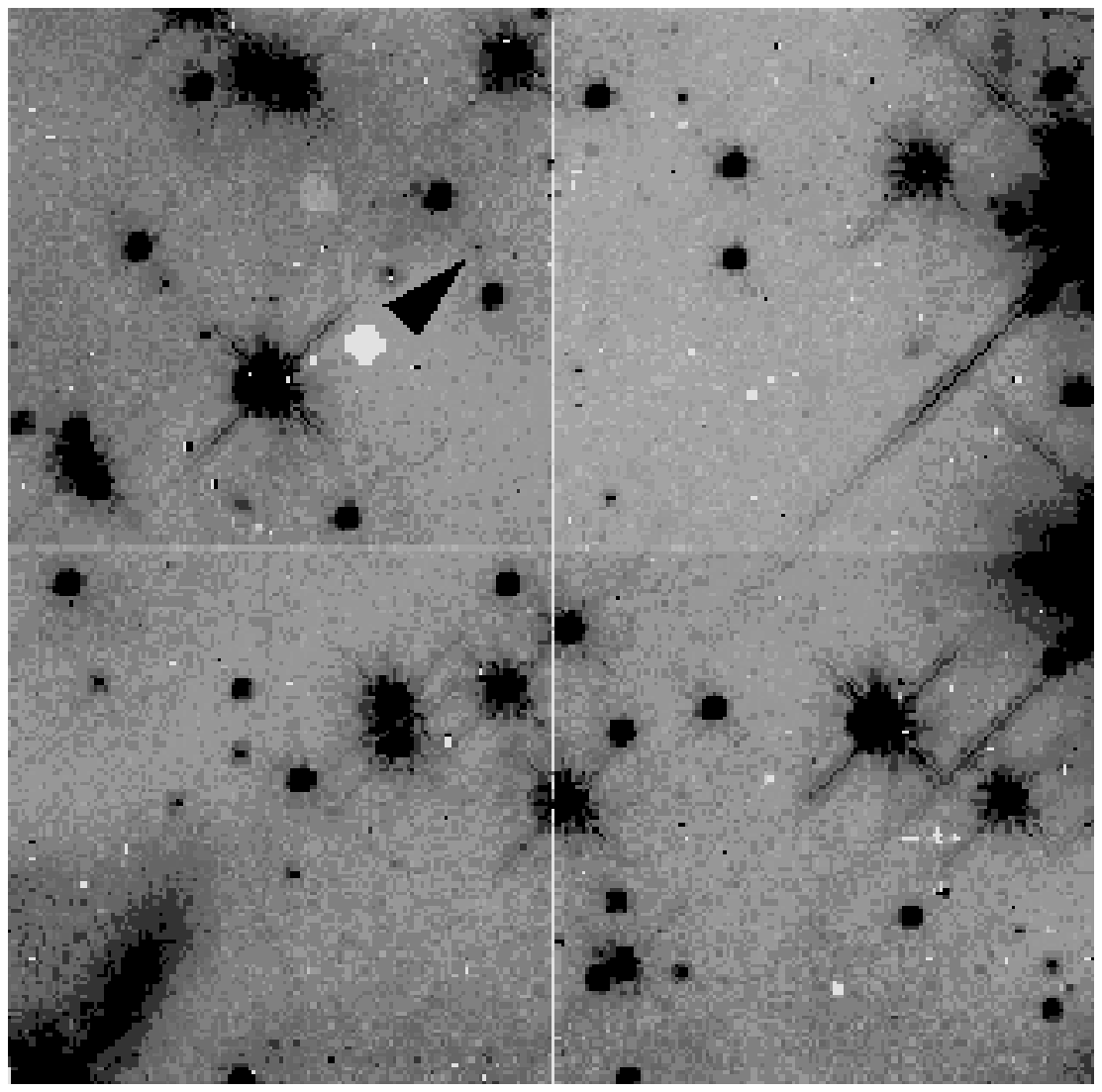}{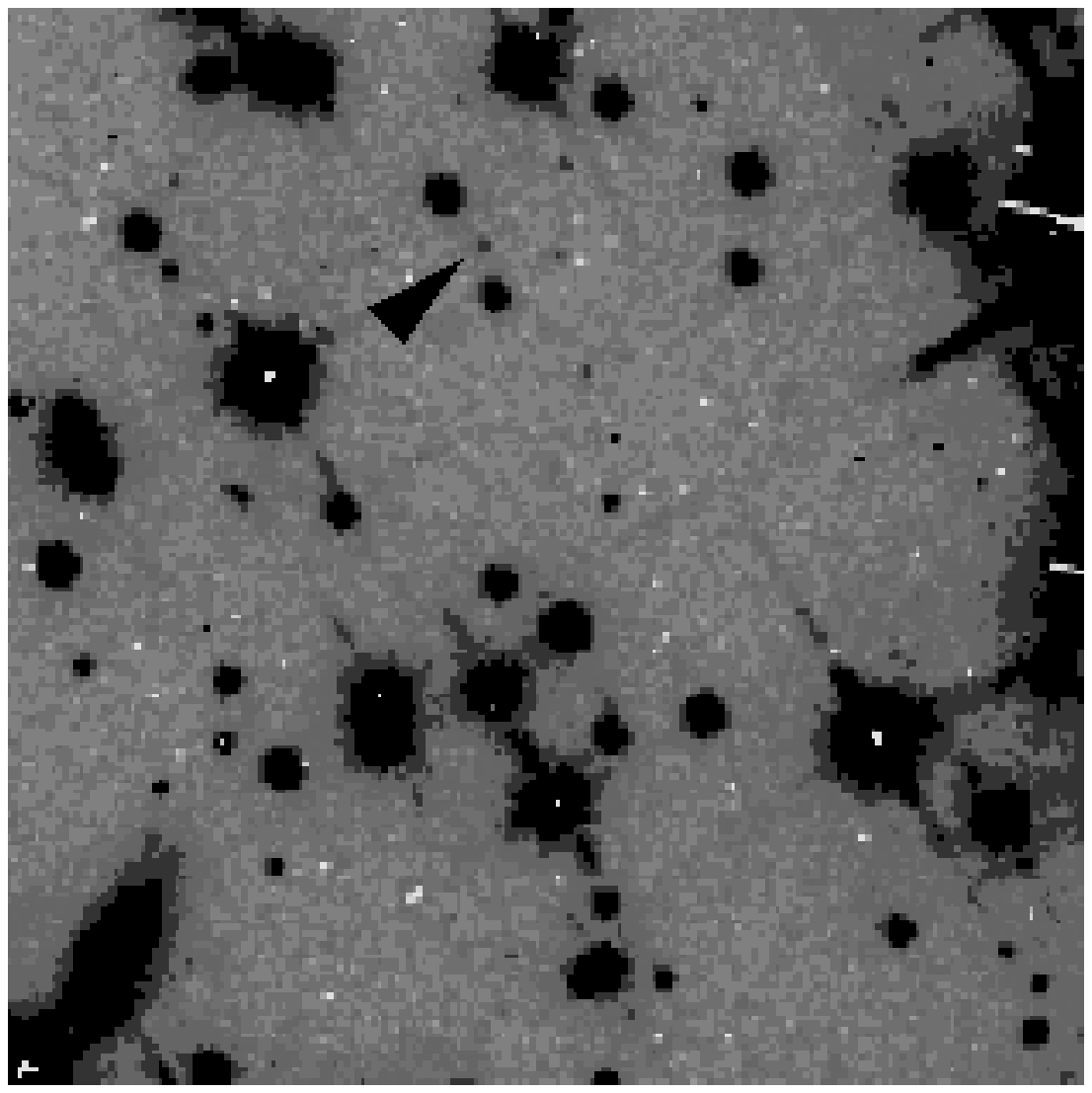}
\caption{Comparison of the same area (field F\,2) as seen in a with
NICMOS NIC\,2 in the $H$ band with a $\sim 170$\,min exposure (left panel)
and in a $\sim 90$\,min duration WFPC\,2 image through the $I$ band
filter (F814W, right panel). The arrow indicates a star close to the
H-burning limit in this cluster (see text).}
\end{figure}

\begin{figure}[h]
\plottwo{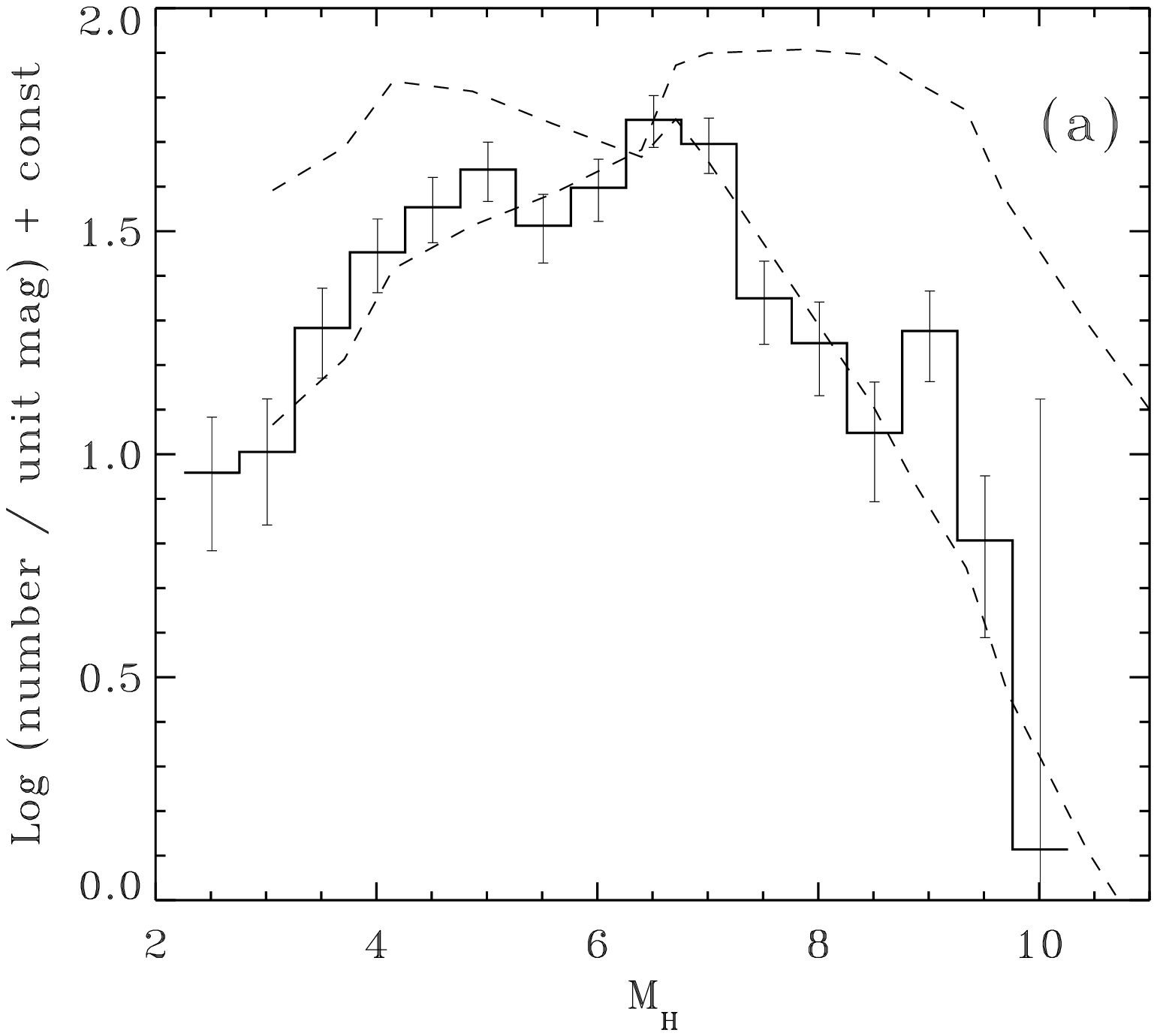}{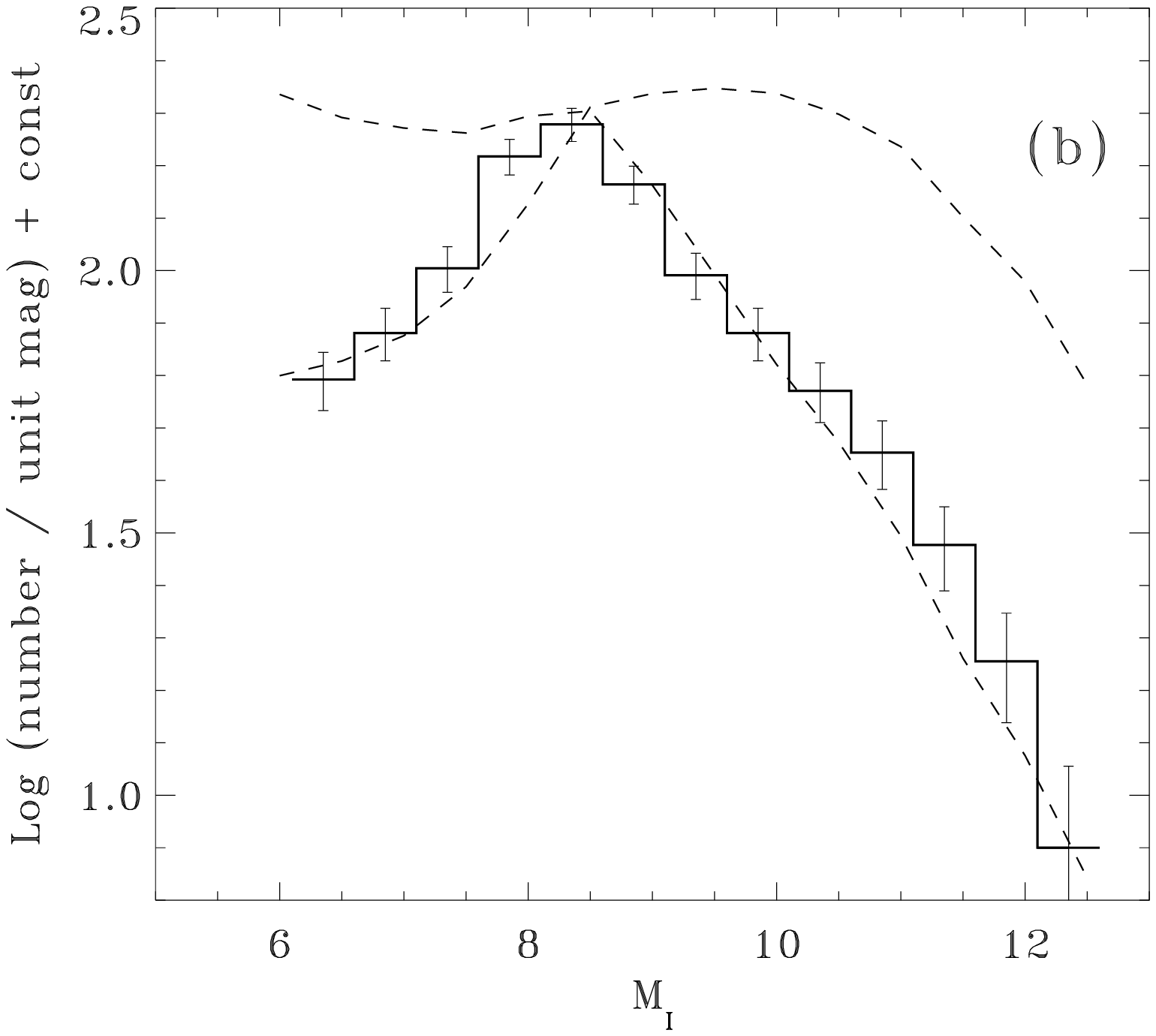}
\caption{Comparison between the IR and optical LF: (a) LF of stars in
field F\,1 in units of absolute $H$-band magnitude $M_{\rm H}$, and (b)
the LF in the $I$ band (F814W) as measured by Paresce et al. (1995)
with the WFPC\,2, also converted to absolute magnitude units $M_{\rm
I}$. The dashed lines show the best fitting underlying MF, characterized 
by $\alpha=1.2$ and $\alpha=-0.6$ respectively for the rising and
dropping portion of the $H$ band LF, and $\alpha = 1.4$ and $\alpha=-0.3$ 
in the $I$ band. Acceptable fits are obtained for $\alpha$ within about 
$\pm 0.2$ of the best fitting values, as explained in the text.}
\end{figure}

NGC\,6397 is the first globular cluster for which a deep, reliable LF
became available from WFPC\,2 photometry of MS stars all the way to the
bottom of the MS (Paresce et al. 1995). So far, two other independent
studies exist with the WFPC\,2 that confirm the deficiency of faint
stars near the half-light radius of this cluster discovered by Paresce
et al. (Cool et al. 1996; King et al.  1998). A direct comparison of
those observations, carried out at optical wavelengths, with our new IR
data could clarify whether the WFPC\,2 is efficient at detecting faint,
red dwarfs or whether it has missed a large amount of the very low mass
objects that could potentially account for a conspicuous fraction of
the total mass of GC.

The first step in this attempt is to compare deep optical ($I$ band)
and IR ($H$ band) images of the same area (field F\,2), as we do in
Figure\,3. It is immediately evident that all the objects detected with
NICMOS in the $H$ band with a $\sim 170$\,min exposure (left panel) are
also present in the $\sim 90$\,min WFPC\,2 image through the $I$ band
filter (F814W, right panel). The arrow indicates a star with $M_V =
13.8$, $M_I = 12.4$, $M_J = 11.4$, and $M_H = 10.7$  which, using the
distance modulus of Reid \& Gizis (1998) and the M-L relation of
Baraffe et al. (1997) for [M/H]$=-1.5$,  translates to a mass of
$\lesssim 0.09$\,\Msolar, namely that expected for an object close to
the H-burning limit in this cluster (Baraffe et al. 1997). Since this
object is clearly visible in all bands ($V$,$I$,$J$, and $H$), it is
unlikely that the WFPC\,2 has underestimated the number of low-mass
stars. In principle, then, one should expect the two instruments to
give the same LF, but since the number of stars common to both NICMOS
and WFPC\,2 is small, statistical fluctuations might be important.

A second, perhaps better way to reveal any significant difference
between the results of NICMOS and the WFPC\,2 data is to compare the IR
and optical LF to one another. The wavelength difference, however,
would require a transformation from one photometric system to the
other, which would inevitably introduce uncertainties that we want to
avoid. To minimize possible systematic errors, we prefer to compare to
one another the theoretical MF that best fit our visible and IR data.

In Figure\,4\,a we show the LF of field F\,1 in units of absolute
$H$-band magnitude $M_{\rm H}$, assuming a distance modulus of
$m-M=12.2$ (Reid \& Gizis 1998) and $E(B-V)=0.18$ (Djorgovski 1993).
Figure\,4\,b shows the LF in the $I$ band (F814W) as measured by
Paresce et al. (1995) with the WFPC\,2, also converted to absolute
magnitude units $M_{\rm I}$. The LF in the $I$ band does not cover the
whole MS, in that it stops at the bright end $\sim 2$ mag before
reaching the turn-off ($M_{\rm I} \simeq 4$) due to saturation.

Since it is often assumed for simplicity that, even over a wide mass
range, the stellar MF is well represented by a power-law distribution
of the form $dN/dm = m^{-\alpha}$ (see, however, Scalo 1998 about the
large range spanned by $\alpha$), we initially adopt such a functional
form for the theoretical MF.\footnote{Some authors prefer to express
the power-law MF in logarithmic units, i.e. $dN/d\log m = m^{-x}$. The
two exponents $\alpha$ and $x$ are such that $\alpha = x + 1$, with
Salpeter's IMF taking on the value of $\alpha = 2.35$ and $x=1.35$,
respectively.  Throughout the rest of the paper we will use the linear
notation ($\alpha$), except for Figure\,6 which, as we explain, is best
displayed in logarithmic units.} The dashed lines in Figure\,4 show the
theoretical LF obtained by multiplying a power-law distribution by the
derivative of the M-L relation appropriate to NGC\,6397 (Baraffe et
al.  1997). This M-L relation is adopted here because it is the only
one presently available that takes into account all the stellar physics
on the problem and that fits very well all available data on the CMD
of globular clusters. The inspection of Figure\,4\,a and 4\,b
immediately reveals that a single-exponent power-law distribution is
not a viable choice for the underlying MF at this position, regardless
of the value of $\alpha$. And indeed, while the bright portion of the
LF up to the peak at $M_{\rm H} \simeq 6$, $M_{\rm I} \simeq 8$
requires $\alpha \simeq 1.4$ (acceptable fits are found for $0.8 <
\alpha < 1.6$ in the $H$ band, and $1.2 < \alpha < 1.6$ in the $I$
band), a much shallower exponent $\alpha \simeq -0.4$ ($-0.8 < \alpha <
-0.4$ in $H$, $-0.5 < \alpha < -0.1$ in $I$) is needed to fit the
dropping portion of the LF beyond the peak. At the same time,
Figure\,4 shows that the same type of MF applies to both the optical
and IR data, i.e. one that flattens out and drops at $\sim
0.3$\,\Msolar, close to the peak of the LF when adopting the M-L
relation of Baraffe et al. with the distance modulus and reddening
values given above. In other words, the peak and turn over observed in
all the LF measured so far for NGC\,6397 near the half-mass radius
(Paresce et al. 1995, De Marchi \& Paresce 1997, King et al. 1998)
cannot be attributed exclusively to the change of slope of the M-L
relation at $\sim 0.3$\,\Msolar.
 
This result is fully consistent with the conclusions of Paresce et al.
(1995) and De Marchi \& Paresce (1997), showing that they apply to the
IR as well. {\it We can, therefore, exclude that near the half-light
radius of NGC\,6397 the MF of the stellar population is represented by
a single-exponent power-law distribution} as advocated by Chabrier \&
M\'era (1997) and Silvestri et al. (1998). Chabrier \& M\'era (1997)
reach this conclusion by assuming that the last four bins of Paresce et
al. (1995) can be safely ignored as being seriously affected by
incompleteness.  This is now known not to be the case.  Silvestri et
al. (1998) reach this conclusion by using a model that is inconsistent
with the observed CMD of NGC\,6397 and other clusters.  This is most
likely due to their use of a grey-like atmosphere approximation which
Chabrier \& Baraffe (1997) have convincingly demonstrated to be wrong
below $\sim 0.6$\,\Msolar.

\section{Local Mass Function or Global Mass Function ?}

It is well known that, through the relaxation process, stars in GC
tend to reach energy equipartition and, as a result, their mass
spectrum can vary in time and space (Spitzer 1987) so that, for
instance, the MF measured in the core of a cluster in thermal
equilibrium should decrease with decreasing luminosity. Paresce, De
Marchi, \& Jedrzejewski (1995),  King, Sosin, \& Cool (1995), and De
Marchi \& Paresce (1996) have indeed shown this to be precisely the
case, in the mass range $0.8 - 0.5$\,\Msolar, respectively for 47\,Tuc,
NGC\,6397, and M\,15. At this juncture, it is fair to wonder whether
the flattening  below $\sim 0.3$\,\Msolar that we now observe in the MF
of field F\,1 and F\,2 is an intrinsic feature of the IMF, or perhaps a
characteristic signature of the global MF, or whether it simply is the
result of some local dynamical effects bearing no relation whatsoever
with either the IMF or the global MF. Although conventional wisdom and
theoretical models suggest that away from both the cluster center and
from its periphery these effects should be rather modest (Richer et
al.  1991; Vesperini \& Heggie 1997), no unambiguous proof exists that
this hypothesis is correct, nor does one exactly know where (if
anywhere) the local MF should be most insensitive to dynamical
modifications.  Thus, since NGC\,6397 is probably in an advanced
evolutionary state and has probably already gone through the collapse
phase, the two-body relaxation process that has strongly changed the
properties of the stellar population in the cluster's core (King et
al.  1995) might have worked its effects farther out in the periphery,
possibly imparting non negligible modifications onto the MF in the
regions that we observed.  And indeed, a careful inspection of
Figure\,4 already reveals that, although similar, the MF that best fits
the LF in field F\,1 is not precisely the same that one requires to
properly reproduce the LF observed in field F\,2 (see discussion
above).

In an attempt to clarify this issue and to put more stringent
constraints on the shape of the cluster's global MF (and, possibly, on
its IMF), we have computed the effect of the segregation mechanism in
NGC\,6397 by using a dynamical model of the cluster so that we could
compare our observations with the predicted shapes of the LF at several
radial positions. Besides the data obtained in field F\,1 ($r \simeq
3\farcm2$) with NIC\,3 and in field F\,2 ($r \simeq 4\farcm5$) with
WFPC\,2 and NIC\,2,  we have included in this exercise the data taken
with the WFPC\,2 by Mould et al. (1996) in a field located $\sim
10^{\prime}$\,N of the cluster center. In order to ensure that the
comparison of these latter data to our own observations is not hampered
by possible differences in the way they were reduced, we have extracted
from the HST archive the images taken by Mould et al.  (1996) and have
subjected them to the same, homogeneous data processing that we have
applied to our WFPC\,2 observations of NGC\,6397 (Paresce et al.
1995).

Observations of Mould et al.'s field (hereafter called F\,3) are
available through both the F555W and F814W bands, with a total exposure
time of $3,600$\,s in each filter. This has allowed us to build a CMD
in all ways similar to that of Mould et al. (see their Figure\,1), to
which we have applied our ``$2.5 \, \sigma$\,-clipping'' criterion to
discriminate, in a statistical way, the stars belonging to the cluster
from field objects (see De Marchi \& Paresce 1995 for a detailed
description of the method). The LF of MS stars determined in this way
agrees reasonably well with that measured by Mould et al., in that both
LF increase with decreasing luminosity up to a peak at $m_{814} \simeq
20.5$ and then fall off. Theirs, however, reveals a secondary peak at
$m_{814} \simeq 22$ ($M_I \simeq 9.5$) that we do not see and which is
most likely the result of a different amount of correction for field
star contamination. We would like to point out that the latter
represents the largest source of uncertainty in the determination of
the LF in field F\,3, and that this effect is much more severe than
that made by ignoring the photometric incompleteness as both Mould et
al. and we have done.  The resulting LF is shown in Figure\,5 after
having been converted into units of absolute magnitude by adopting the
distance modulus given above.

\section{Modelling Mass Segregation}

\begin{figure}[h]
\plotone{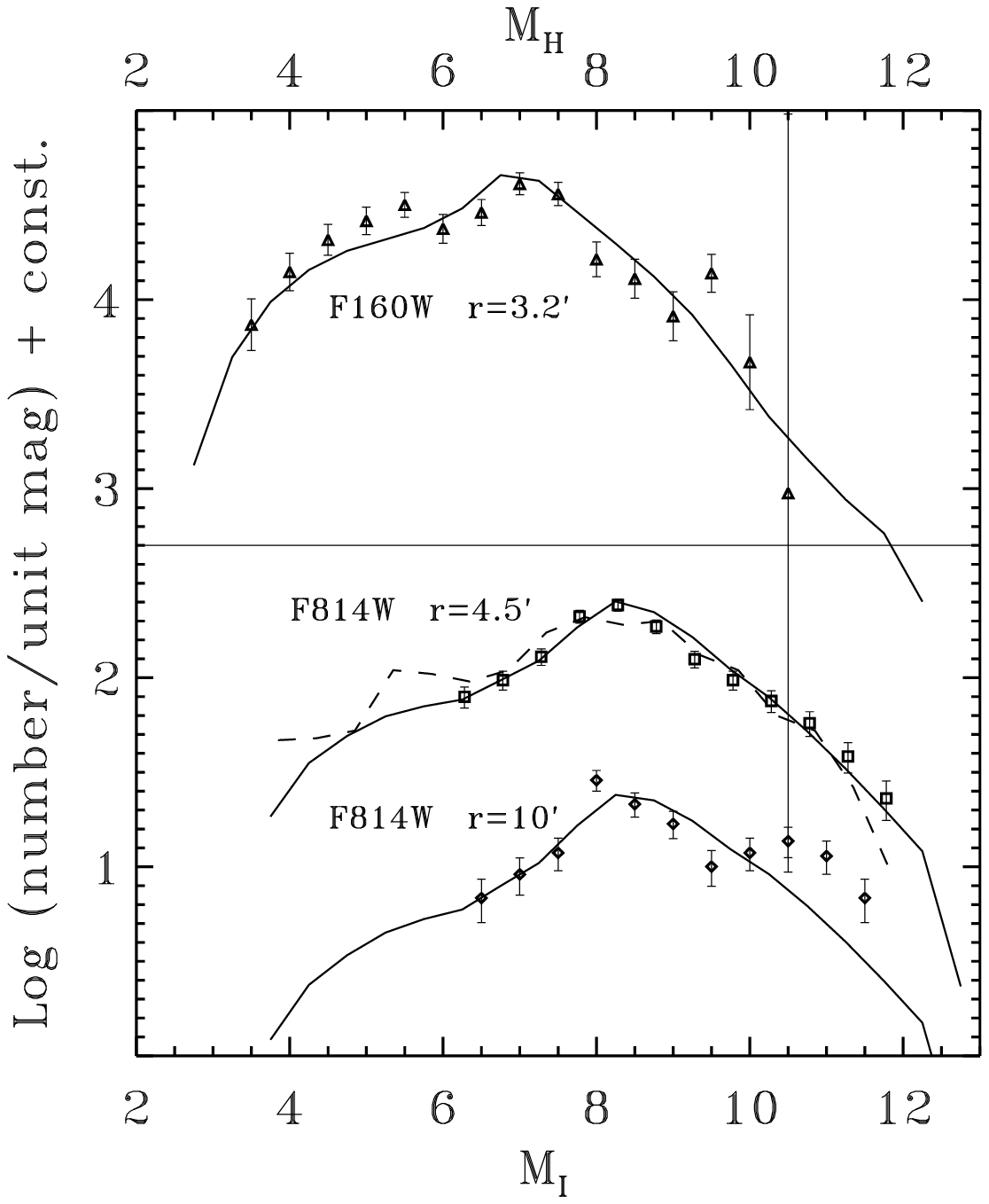}
\caption{The solid lines show the LF predicted by the dynamical model
that best fits simultaneously the data at $3\farcm2$ (triangles),
$4\farcm5$ (boxes), and $10^{\prime}$ (diamonds) radial distance as
well as the SBP and VDP (see text).  The scale on the top axis refers
only to the IR LF at $r=3\farcm2$. The dashed line shows the LF as
measured by King et al. (1998) in a different region of the cluster,
yet still $\sim 4\farcm5$ away from the center.}
\end{figure}

We have employed the multi-mass Michie-King models developed by Meylan
(1987, 1988), and which were later adopted by Meylan \& Mayor (1991) to
study the dynamical properties of NGC\,6397. Each model is
characterized by an IMF in the form of an exponential  $dN/dm =
m^{-\alpha}$, with a variable exponent $\alpha$, and by four structural
parameters describing respectively the scale radius ($ r_{\rm c}$), the
scale velocity ($v_{\rm s}$), the central value of the dimensionless
gravitational potential $W_{\rm o}$, and the anisotropy radius ($r_{\rm
a}$). From the parameter space defined in this way, Meylan \& Mayor
(1991) have selected those models that simultaneously fit both the
observed surface brightness (SBP) and velocity dispersion (VDP)
profiles of the cluster. The fit to the SBP and VDP, however, can only
constrain $r_{\rm c}$, $v_{\rm s}$, $W_{\rm o}$, and $r_{\rm a}$ while
still allowing the IMF to take on a variety of shapes. To break this
degeneracy, we have complemented Meylan's code by imposing the
condition that the model MF agree with the observed LF, following
precisely the approach that we used in our investigation of the
globular cluster M\,4 (Pulone, De~Marchi, \& Paresce 1999).

As we explain in that paper, Michie-King modeling only provides a
``snapshot'' of the current dynamical state of the cluster.  It is,
then,  useful to define the global mass function (GMF), i.e.the mass
distribution of all cluster stars at present, as the MF that the
cluster would have simply as a result of stellar evolution (i.e.
ignoring any local modifications induced by internal dynamics and/or
the interaction with the Galactic tidal field). Clearly, in this case
the IMF and GMF of MS (un-evolved) stars is the same. For practical
purposes, the GMF has been divided into sixteen different mass classes,
covering MS stars, white dwarfs, and heavy remnants, precisely as
described in Pulone et al.  (1999).

The SBP and VDP used in our simulations are taken from the paper of
Meylan \& Mayor (1991), as G. Meylan kindly gave us the data in
electronic form. Since these authors provide the mass, density, and
structural parameters of the 8 models that best reproduce the observed
radial profiles (SBP and VDP), we have tested whether the predicted LF
would also agree with our observations.  Our exercise confirms what we
have already shown in Figure\,4: as long as a single value of the
exponent $\alpha$ is used for the IMF over the mass range $0.1 -
0.8$\,\Msolar, none of these 8 models gives local MF that, converted
into LF using the M-L relation described above, agree with our data.
In fact, if the exponent $\alpha$ is selected in such a way that the
rising portion of the LF is well fitted (this requires $\alpha \simeq
1.6$), the model LF consistently deviate from the observations beyond
the peak at $M_I \simeq 8.5$, $M_H \simeq 6.5$ because they fall off
too slowly.  Similarly, to fit the falling portion of the LF one would
require $\alpha \simeq - 0.2$, which would in turn overestimate the
number of bright stars.

On the other hand, if the model IMF is allowed to flatten out below
$\sim 0.3$\,\Msolar, both portions of the three observed LF can
simultaneously be fitted while, at the same time, the SBP and VDP are
adequately reproduced. In particular, we find that an exponential IMF
with index $\alpha = 2 \pm 0.2 $ for evolved stars ($ m > 0.8$\,\Msolar), $\alpha = 1.6 \pm 0.2$ in the range $0.8\mbox{--}0.3$\,\Msolar and
which drops with $\alpha = -0.2 \pm 0.1$ at lower masses would not only
produce a good fit to the LF observed in field F\,1, F\,2, and F\,3
(Figure\,5), but would also give reasonable values for the cluster's
structural parameters:  the concentration ratio ($c \simeq 2.6$), the
total cluster mass ($\simeq 8.9 \times 10^4$\,\Msolar), the fraction of
heavy remnants ($\sim 33\,\%$ in mass, about half of which in the form
of neutron stars of $1.2$\,\Msolar and the rest in the form of heavy
white dwarfs of $0.6$\,\Msolar), and the anisotropy radius ($r_{\rm a}
/ r_{\rm c} \simeq 20$) are all in excellent agreement with those of
the 8 best models of Meylan \& Mayor (1991).

King et al. (1998) notice that the somewhat large value of the core
radius that Meylan \& Mayor (1991) obtain (and that we only marginally
decrease here) does not allow them to accurately reproduce the SBP in
the innermost regions of the cluster. On the other hand, Meylan \&
Mayor point out that the deviation of the observed profile from that of
a typical Michie-King model in the central regions is most likely the
result of random fluctuations due to a few bright stars, whose
distribution is not uniform (Auri\`ere, Lauzeral, \& Ortolani 1990).
Moreover, if the core of NGC\,6397 has already been through its collapse
phase, a classical King-type profile (King 1962) of a cluster in
equilibrium certainly does not provide the most appropriate
representation of the SBP and the value of $r_{\rm c}$ that one would
deduce from it would be rather uncertain.

To test whether our model also gives a proper account of the properties
of the stellar population in the core, we have compared the MF that it
predicts at $\sim 7$\arcsec from the center with that measured by King
et al. (1995) with the FOC on board HST. Although our model seems to
predict a somewhat slower decline than that reported by King et al. in
the range $0.7-0.4$\,\Msolar (namely $\alpha \simeq -2$ vs.  their
$\alpha \simeq -2.7$), this discrepancy should not be taken too
seriously in view of the limitations of our model and of the
uncertainties that could affect the conversion of the observed LF from
the F480LP band of the FOC into the Johnson's V band, and then from
that into the F814W band of the WFPC\,2 and, finally, into a MF by the
use of a rather crude M-L relation.  We can, therefore, conclude that
the two MF are in fairly reasonable agreement with each other, in that
they both indicate an inverted function (i.e. stars are decreasing with
decreasing mass) starting already at the turn-off.

\section{Discussion}

\begin{figure}[h]
\plotfiddle{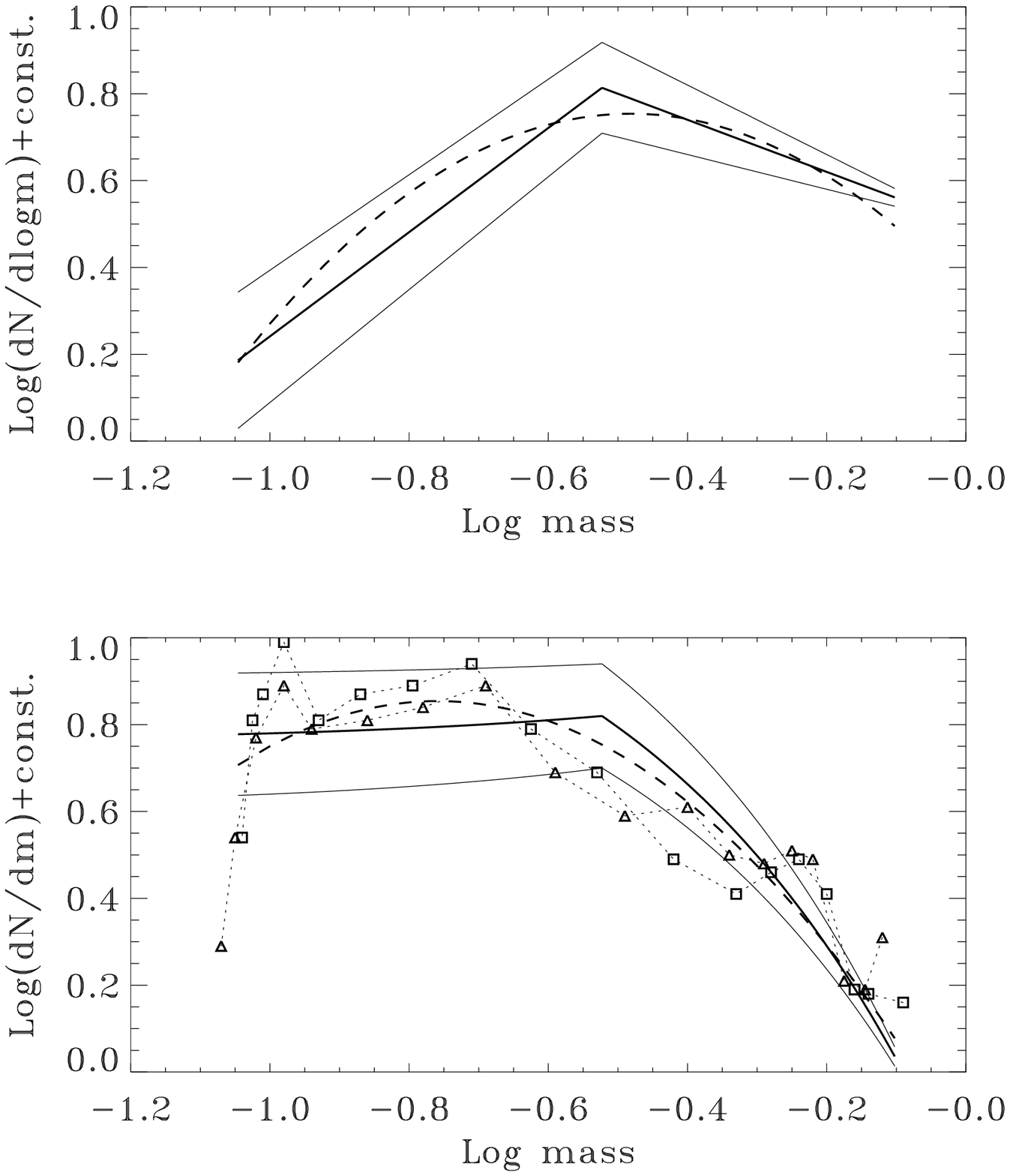}{9cm}{0}{55}{55}{-180}{-80}
\caption{Top panel: The GMF of NGC\,6397 as predicted by our dynamical
model (thick solid line) is consistent, within the error band marked by
the thin lines, with a log-normal distribution having $m_{\rm c} \simeq
0.3$\,\Msolar and $\sigma \simeq 1.8$ (dashed line). The error band
corresponds to the uncertainty in the slopes of the two power-law
segments used to represent the IMF in our dynamical model (see
Sections\,3 and 5) and, as such, it does not directly reflect the
statistical fluctuations of each individual bin in our LF. Please note
that, because the log-normal distribution is best represented in units
of logarithmic mass, also the power-law curves in the figure are
plotted in the same units, i.e. $dN/d\log m = m^{-x}$. The exponent
$\alpha$ used throughout the text obeys the relation $\alpha=x+1$ as
explained in Section\,3.
Bottom panel: Same GMF as above, but in linear mass units, i.e.
$dN/dm$. The dotted lines represent the MF of NGC\,6397 as obtained by
King et al. (1998) using the mass--luminosity relations of Baraffe et
al. (1997), respectively for a distance modulus of 12.05 (boxes) and
12.05 (triangles). See King et al. (1998) for the details.
}
\end{figure}

The result stemming from Figure\,5 and from the discussion above is
fully consistent with our findings in Section\,3.  {\it Not only can we
exclude that the local MF measured near the half-light radius come
from a single exponent power-law, but we also prove here that not even
the global MF from which they originated through energy equipartition
can be a function of that type}.  These findings carry the important
implication that, if NGC\,6397 has not experienced too strong an
interaction with the Galactic tidal field or bulge (but see Gnedin \&
Ostriker 1997), also the IMF has to flatten out and, possibly, drop
below $\sim 0.3$\,\Msolar.

Since, however, Nature has always shown a predilection for smooth
variations, it might seem rather unphysical that the IMF or GMF of MS
stars can have such an abrupt change of slope at $\sim 0.3$\,\Msolar as
our work seems to imply. Clearly, our fit to the GMF is nothing but a
mathematical over-simplification, dictated by the adoption of a
power-law as representative of the distribution of stellar masses. As
Scalo (1998) points out, this assumption might be correct for
intermediate mass stars, but fails at the edges of the distribution
(very low and very high masses). Although we could have used more than
two values of the exponent $\alpha$ to fit the change of slope, that
would have appeared as an unwarranted sophistication  whose physical
meaning remains uncertain.  Thus, since the most important implication
of Figures\,4 and 5 is that the MF of NGC\,6397 cannot be reproduced
using a single exponent power-law distribution, one wonders why
choosing two or more power-laws would make any sense at all.

There is, however, a large amount of theoretical work suggesting that
the star formation process would preferentially give rise to a mass
distribution that deviates from a pure power-law. Zinnecker (1984) and,
more recently, Adams \& Fatuzzo (1996) have pointed out that, if the
number of parameters governing the fragmentation process of
proto-stellar clouds is sufficiently large (of order 5 or more), the
central-limit theorem immediately implies that  the IMF should take on
a log-normal form. A similar IMF is obtained independently by Elmegreen
(1997, 1999) with a different theoretical approach. If we adopt the
formalism of Adams \& Fatuzzo (1996), the IMF is characterized by two
parameters, namely the characteristic mass $m_{\rm c}$ and the standard
deviation $\sigma$ and takes on the form:

\begin{equation}
\ln f \left( \log m \right) = A - \frac{\left[ \log(m/m_c)\right]^2}
{2 \sigma^2 }
\end{equation}

where $A$ is a normalization constant. The Miller \& Scalo (1979) IMF
(now superseded; Scalo 1998) would require  $m_{\rm c} \simeq
0.095$\,\Msolar and $\sigma = 1.57$, while the Scalo (1986) IMF (now
also extinct; Scalo 1998) would imply both a larger value of $m_{\rm
c}$ and a smaller width $\sigma$. In Figure\,6 we show that, within the
observational uncertainties, the GMF of NGC\,6397 is consistent with a
log-normal distribution having $m_{\rm c} \simeq 0.3$\,\Msolar and
$\sigma \simeq 1.8$. Interestingly, these values are very similar to
those found for MS stars in the Pleiades (Bouvier et al. 1998).
  
\section{Summary and Conclusion}

The main results of this paper can be summarized as follows:

\begin{itemize}

\item 
The properties of the stellar population of NGC\,6397 as determined
with deep IR observations obtained with NICMOS on board the HST are
fully consistent with those stemming from similarly deep WFPC\,2
optical data. In particular, down to the bottom of the cluster's MS,
NICMOS does not show any more faint, red objects than those already
detected by the WFPC\,2, thus confirming that there are no stars
contributing significantly to the low-mass budget of the cluster which
can be detected in the IR and not in the $I$ band. Thus, the
unavailability of NICMOS will not hinder this investigation in the
future.

\item
A comparison of our deep IR CMD with the theoretical models of Baraffe
et al. (1997) shows the agreement to be excellent, thus proving that
the latter are reliable also in the IR domain.

\item
The IR LF of MS stars near the cluster's half-light radius, measured
using a comparison field located a few degrees away from the cluster
center to correct for the contamination due to foreground and
background objects, is fully consistent with those obtained with the
WFPC\,2 in the $I$ band using different and independent techniques to
separate cluster members from field stars, namely the color selection
of Paresce et al. (1995) and the proper motion study of King et al.
(1998).

\item
Using the M-L relations of Baraffe et al. (1997), all three LF
translate into MF in which the number of objects per unit mass
increases exponentially as $\sim m^{-1.5}$ in the range $0.7 -
0.3$\,\Msolar and then flattens out and drops. We show that a simple
power-law distribution of the form $m^{-\alpha}$ cannot reproduce any
of these LF if the exponent $\alpha$ is kept constant, regardless of
its value.

\item
The relaxation mechanism and the ensuing mass segregation process
inside the cluster is studied under the assumption that a simple
isothermal model of a cluster in equilibrium is a viable choice for
NGC\,6397 and ignoring the effects of tidal stripping due to the
Galaxy, as they do not seem to play any significant role in reshaping
the MF of this cluster (Paresce \& De Marchi 1999). The model can
accurately and simultaneously reproduce the LF observed in regions
located $\sim 3\farcm2$, $4\farcm5$, and $10^{\prime}$ away from the
center, the inverted LF in the core, as well as the SBP and VDP of red
giant stars only if the IMF rises as $m^{-1.6 \pm 0.2}$ in the range
$0.8 - 0.3$\,\Msolar and then drops as $m^{0.2 \pm 0.1}$ below $\sim
0.3$\,\Msolar. The error bands associated with the quoted exponents
assume that the M-L relations of Baraffe et al. (1997) are not affected
by statistical uncertainties.

\item
The IMF that we obtain is fully consistent with a log-normal
distribution peaked at $m_{c} \simeq 0.3$\,\Msolar and with standard
deviation $\sigma \simeq 1.8$.

\end{itemize}

In this paper we show that a robust and reliable determination of the
IMF of GC requires the effects of dynamical evolution to be taken into
proper account. Although we have used the best LF currently available
for NGC\,6397, we are still forced to rely on uncertain brightness and
velocity dispersion radial profiles and on very simple Michie--King
multi-mass quasi-equilibrium models to convert a locally determined
present day mass function into a global MF. In fact, while the models
that we have used are currently the state of the art, they work only
under the strict condition that the cluster be fully relaxed and in
dynamical equilibrium. Episodes of core collapse and, especially,
gravitational shocking due to repeated interactions with the bulge and
disk of the Galaxy, however, may profoundly disrupt this distribution
by ejecting low mass stars from the core and by compressing the tidal
boundary in phase space at each encounter (see e.g. De Marchi et al.
1999). Until these effects can properly be accounted for in future
models of this sort, we should obtain the global cluster MF directly by
observation through a systematic, precise, and complete census of the
entire stellar population of  GC spanning a wide range of possible
evolutionary scenarios. Specifically, one needs to determine the exact
shape of the cluster LF at many locations in the cluster from the main
sequence turn-off down to at least the peak of the LF at $\sim
0.3$\,\Msolar.  These local LF and the resulting global MF can then be
compared to the available cluster models to assess their validity under
differing circumstances. This project is within easy reach of a large
telescope such as the VLT.

We would like to thank Georges Meylan for providing us with the
original code of his dynamical model and Isabelle Baraffe and Gilles
Chabrier for many useful discussions.\\
We are indebted to Haldan Cohn, the referee of this work, for his
invaluable comments that have considerably strengthened our paper.

\end{document}